\documentclass[12pt]{article}
\usepackage[margin=1in]{geometry}
\usepackage{graphicx,url}

\usepackage[utf8]{inputenc}  

\usepackage{authblk} 

\usepackage{amsmath,amsfonts,amssymb}
\usepackage{hyperref}
\usepackage{bm} 

\usepackage{pdfpages}
\usepackage{natbib}
    
\title{On community structure validation in real networks}

\author[1]{Mirko Signorelli$^*$}
\author[2]{Luisa Cutillo\footnote{The authors contributed equally to this paper.}}

\affil[1]{Mathematical Institute, Leiden University (NL)}
\affil[2]{School of Mathematics, University of Leeds (UK)}

\date{}

\begin{document} 

\maketitle

\vspace{-1cm}

\noindent \textbf{About this article}
\begin{itemize}
\item Please cite this article as: Signorelli, M., Cutillo, L. (2021). On community structure validation in real networks. \textit{Computational Statistics}. DOI:\\10.1007/s00180-021-01156-6.
\item This document contains the ``accepted'' version of the manuscript. The final (published) version of the article can be freely downloaded (Open Access) from the website of \textit{Computational Statistics}, using this link: \href{https://doi.org/10.1007/s00180-021-01156-6}{https://doi.org/10.1007/s00180-021-01156-6}
\end{itemize}

\begin{abstract}

\textit{Community structure} is a commonly observed feature of real networks. The term refers to the presence in a network of groups of nodes (\textit{communities}) that feature high internal connectivity, but are poorly connected between each other.
Whereas the issue of community detection has been addressed in several works, the problem of validating a partition of nodes as a good community structure for a real network has received considerably less attention and remains an open issue.
We propose a set of indices for community structure validation of network partitions that are based on an hypothesis testing procedure that assesses the distribution of links between and within communities.
Using both simulations and real data, we illustrate how the proposed indices can be employed to compare the adequacy of different partitions of nodes as community structures in a given network, to assess whether two networks share the same or similar community structures, and to evaluate the performance of different network clustering algorithms.

\hspace{-0.1cm}\\
\noindent \textbf{Keywords:} community structure; community validation; graph; network clustering; network  enrichment analysis; stochastic blockmodel.

\end{abstract}

\section{Introduction}\label{sec:intro}

The growing availability of data on real world networks has inspired the study of complex networks in the multidisciplinary fields of social, technological and biological networks. What makes networks so attractive? We are constantly dealing with networks: supermarkets use networks to propose specific deals to targeted groups; banks orchestrate a complex system of transactions between them and clients; media networks dominate our lives, and inside each living being genes express and co-regulate themselves via complex networks (even when we sleep). Graphs constitute a mathematical representation of complex systems, whose understanding requires a careful study of their structure. Such a task can be particularly challenging for large graphs featuring hundreds or thousands of nodes.

The study of the structure of a graph is often achieved by decomposing it into its constituent modules or communities. \citet{Girvan2002} address the concept of community structure as a network property. Indeed networked systems can be described via main statistical properties such as small-world property, power-law degree distributions, network transitivity and clustering coefficient. They highlight that the property of community structure is found in most real networks. This essentially means that nodes within a network are connected together in tightly joined groups, while between those groups connections are looser. 
Identifying communities in a network is highly relevant, as it enables to disclose the presence of an internal network structure at a very preliminary analysis step. Over the years, a significant effort has been devoted to the development of several community detection algorithms \citep{clauset2004,pons2005,newman2006,blondel2008,newman2016}, with a strong focus on the scalability of these methods to large networks. 

In applied network analysis, the communities that constitute a network are usually unknown. A network may not have any property of community structure; or, even if it does have a community structure, the communities remain unknown and have to be reconstructed via a community detection algorithm. Once the communities have been estimated with a network clustering algorithm, the analyst is then left with questions on the adequacy of the retrieved clusters. On the one side, the nodes may be misclassified (assigned to the wrong community); on the other, the graph at hand may not have a true underlying community structure, and the clusters may thus be scarcely relevant. The question that motivates our work is thus: how can we evaluate when a partition of a given network is \textit{meaningful}? 

In the analysis of real networks where no information on the true communities is available, it is common practice to try to relate the clusters obtained through a certain community detection algorithm to known features of the nodes. If an association between some features and the clusters can be found, this may be taken as a confirmation of the goodness of the clusters. However, this practice is problematic, because it does not take into account network topology, which is what actually determines the community structure. For example, a network may exhibit a strong community structure even if this structure cannot be related to any observed feature of the nodes; and, on the contrary, it is possible that in a network without community structure, a community detection procedure may produce clusters that can be related to certain nodal attributes, but that are nevertheless meaningless. Thus, a more robust approach to community validation, based on network topology, is needed.

Two recent attempts to assess the quality of network partitions have focused on the possibility to improve clustering methods by exploiting metadata, i.e. additional information about the nodes \citep{newman2016, peel2017}. \citet{newman2016} proposed a novel clustering method that combines a network and its metadata, arguing that relevant metadata can improve the performance of clustering methods. On the other hand, \citet{peel2017} argued against the use of metadata as ground truth for the assessment of the quality of clusters in real-world network. Finally, \citet{carissimo2018} recently proposed a method for the evaluation of network partitions that does not take metadata as ground truth; instead, the method evaluates the stability of the partition recovered by a given algorithm against random perturbations of the original graph structure.

\begin{figure*}
\begin{center}
	\includegraphics[scale = 0.5]{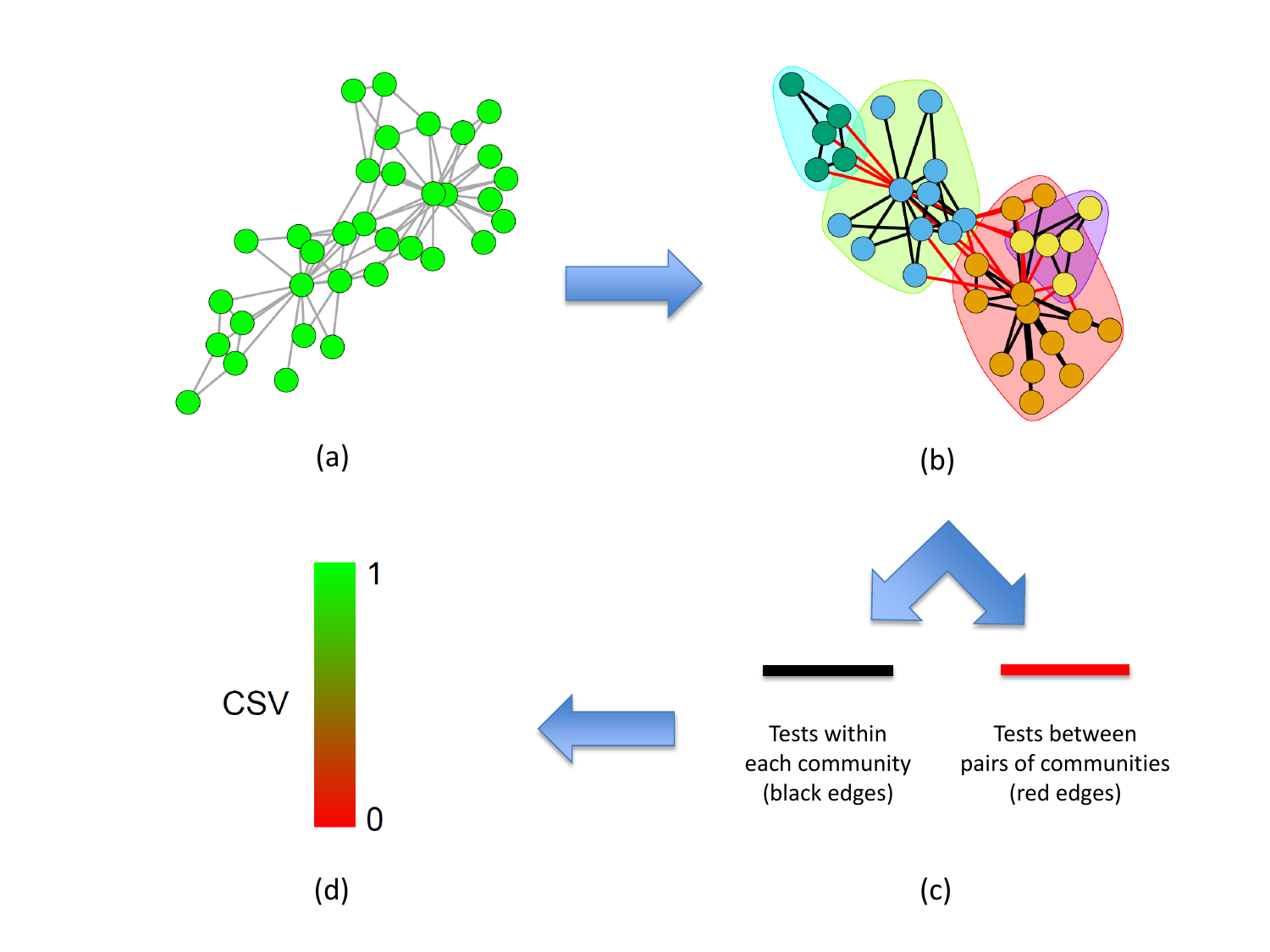}
\end{center}
	\caption{\label{fig:idea} Graphical representation of the construction of the Community Structure Validation index. (a) Consider a binary graph of interest. In the example, we use the \textit{Zachary Karate club} graph. (b) Define a partition of the nodes into $q$ clusters or communities. The communities may either be known a priori, or be the result of a clustering procedure. (c) Following the testing procedure outlined in Section \ref{sub:procedure}, perform $q$ tests of enrichment within each community and $q(q-1)/2$ (or $q^2-q$ if the graph is directed) tests of enrichment between each pair of different communities. (d) Combine the results of the tests thus performed in the Community Structure Validation Index (CSV), as illustrated in Section \ref{sub:CSV}.}
\end{figure*}

In this article we propose a new method for the validation of network partitions as community structures. We focus on the idea that the validation of network partitions should primarily focus on the distribution of links between nodes in the different clusters, rather than simply consider the distribution of nodes among clusters or try to identify observed attributes that correlate with the clusters. Indeed, when assessing the goodness of different partitions of a network, intuitively we would like to rate better partitions with more links connecting nodes within the same community, and less links connecting nodes in different communities. Our method is based on a significance testing procedure for the number of links that are observed between and within the communities; the results from these tests are then combined into a community structure validation (CSV) index that provides an overall assessment of whether a certain partition of nodes induces a community structure in the network. Figure \ref{fig:idea} summarizes the steps for the construction of the CSV index.

Our work borrows the concept of network enrichment from the literature on cross-talk enrichment between gene sets and pathways in biological networks, implementing in particular a one-tailed adaptation of the Network Enrichment Analysis Test (NEAT) proposed by \citet{signorelli2016}. Although the comparison of genetic networks has been an important driver of our work, we emphasize that the proposed methodology is more general and it can be applied to other types of networks as well. Our approach provides also a practical way of comparing networks, by assessing similarity and differences in their community structures.

The remainder of the paper is organised as follows: in Sections \ref{sec:methods} and \ref{sec:netw-comp} we describe the construction of community structure validation indices and we discuss how they can be employed to validate network partitions and to compare community structures across networks. The proposed methodology is evaluated through simulations in Section \ref{sec:simulations} and illustrated with two example applications in Section \ref{sec:application}. Section \ref{sec:concl} briefly summarizes the results obtained in this paper.

\section{Community structure validation}
\label{sec:methods}

In this Section we introduce the main methodological contributions of our work. In Section \ref{sub:procedure} we propose a statistical testing procedure to assess whether a specific partition is a valid community structure for a given graph. In Section \ref{sub:CSV} we define a set of indices that summarize the results from the testing procedure, allowing to quantify the strength of the evidence that a partition induces a valid community structure in a network.\\
We have implemented the proposed methodology in the \texttt{R} programming language \citep{Rlang}. The code used in this paper is available from \href{https://github.com/mirkosignorelli/csv}{\texttt{github.com/mirkosignorelli/csv}}.

\subsection{Inferential procedure}\label{sub:procedure}

We consider a graph $\mathcal{G} = (V, E)$, which consists of a set of vertices (or nodes) $V$ connected by a set $E$ of edges or arrows. Although in this paper we focus mostly on undirected graphs, our approach can be applied to directed and mixed graphs as well. We denote by $\mathcal{P}_V = \{ C_1,...,C_q \}$ a partition of $V$ into $q$ disjoint sets, such that $C_r \cap C_s = \emptyset$ if $r \neq s$ and $\cup_{r=1}^q C_r = V$.

In order to assess whether $\mathcal{P}_V$ induces a community structure in $\mathcal{G}$, we compare the observed number of links within and between each set with the number of links that we would expect to observe by chance if the groups were irrelevant. We do this by implementing a one-tailed adaptation of NEAT, the Network Enrichment Analysis Test proposed by \citet{signorelli2016}. For undirected networks, NET compares the observed number of edges $n_{AB}$ between the set of nodes $A$ and $B$ with an hypergeometric null model which assumes that
\begin{equation}
N_{AB} \sim \text{hypergeom} \left( n = d_A, K = d_B, N = d_V \right),
\label{neat-undirected}
\end{equation}
where $d_A$, $d_B$ and $d_V$ denote the total degrees of sets $A$, $B$ and $V$. For directed networks, NEAT compares the observed number of arrows $n_{AB}$ from the set of nodes $A$ to the set of nodes $B$ with
\begin{equation}
N_{AB} \sim \text{hypergeom} \left( n = o_A, K = i_B, N = i_V \right),
\label{neat-directed}
\end{equation}
where $o_A$ denotes the outdegree of $A$ and $i_B$ and $i_V$ are the indegrees of $B$ and $V$.

In its original implementation \citep{signorelli2016}, NEAT tests the null hypothesis $H_0: \: \mu_{AB} = \mu_{AB}^0$ that the expected number of edges (arrows) between $A$ and $B$,  $\mu_{AB} = E(N_{AB})$, is equal to the expected number of links $\mu_{AB}^0 = E(N_{AB} | H_0) = nK/N$ obtained from models \eqref{neat-undirected} or \eqref{neat-directed} against the two-tailed alternative $H_1: \: \mu_{AB} \neq \mu_{AB}^0$. Here, instead, we implement a one-tailed adaptation of NEAT and consider two distinct one-tailed tests, one for overenrichment, $H_0: \: \mu_{AB} = \mu_{AB}^0$ vs $H_1: \: \mu_{AB} > \mu_{AB}^0$, and one for underenrichment, $H_0: \: \mu_{AB} = \mu_{AB}^0$ vs $H_1: \: \mu_{AB} < \mu_{AB}^0$.

Since a community structure features high internal connectivity within each community and few connections between different communities, we assess the extent to which $\mathcal{P}_V$ generates a community structure by testing
\begin{enumerate}
\item overenrichment within each community $C_r$, $r \in \{1,...,q\}$:
\begin{equation}
H_0: \:\: \mu_{rr} = \mu_{rr}^0, \:\: H_1: \mu_{rr} > \mu_{rr}^0,
\label{overenr-within-test}
\end{equation}
where $\mu_{rr} = E(N_{rr})$ denotes the expected number of links between nodes in $C_r$, and $\mu_{rr}^0$ is the corresponding null expectation from model \eqref{neat-undirected} if $\mathcal{G}$ is undirected, or from model \eqref{neat-directed} if it is directed. Then, we compute the mid-p-values $p_{rr} = \frac{1}{2} P\left(N_{rr} = n_{rr}|H_0\right) + P\left(N_{rr} > n_{rr}|H_0\right)$;
\item underenrichment between each pair of communities $(C_r, C_s)$, with $r < s \in \{1,...,q\}$ if $\mathcal{G}$ is undirected or $r \neq s$ if it is directed:
\begin{equation}
H_0: \:\: \mu_{rs} = \mu_{rs}^0, \:\: H_1: \mu_{rs} < \mu_{rs}^0,
\label{underenr-between-test}
\end{equation}
where $\mu_{rs} = E(N_{rs})$ denotes the expected number of links between nodes in sets $C_r$ and $C_s$, and $\mu_{rs}^0$ the null expectation from model \eqref{neat-undirected} if $\mathcal{G}$ is undirected, or \eqref{neat-directed} if it is directed. Here, we obtain the mid-p-values $p_{rs} = \frac{1}{2} P\left(N_{rs} = n_{rs}|H_0\right) + P\left(N_{rs} < n_{rs}|H_0\right)$.
\end{enumerate}

Because the procedure outlined above requires the computation of $q(q+1)/2$ tests for undirected graphs, or $q^2$ tests for directed graphs, we account for multiple testing using the multiple testing correction procedure proposed by \citet{heyse2011} (which is an adaptation of the Benjamini--Hockberg method for p-values obtained from discrete test statistics) and we derive the adjusted p-values $\tilde{p}_{rr}$ and $\tilde{p}_{rs}$. Alternative multiple testing corrections, or no multiple testing correction at all, may be considered as well.

\subsection{Community structure validation indices} \label{sub:CSV}

Ideally, evidence that a partition induces a clear community structure is strongest if every null hypothesis is rejected for a given type I error $\alpha$, i.e., $\tilde{p}_{rr} < \alpha$ and $\tilde{p}_{rs} < \alpha$ $\forall r,s$. More generally, a large proportion of rejections can be regarded as sufficient evidence of a valid community structure. We summarize this intuition through a set of Community Structure Validation (CSV) indices. The steps involved in the construction of the CSV indices are illustrated in Figure \ref{fig:idea}.

\noindent For undirected graphs, we define the CSV index as
\begin{equation}
CSV_U = \frac{\sum_{r=1}^q I(\tilde{p}_{rr} \leq \alpha) + \sum_{r>s} I(\tilde{p}_{rs} \leq \alpha)}{q(q+1)/2}.
\label{ucsv-undir}
\end{equation}
The corresponding index for directed graphs is given by
\begin{equation}
CSV_D = \frac{\sum_{r=1}^q I(\tilde{p}_{rr} \leq \alpha) + \sum_{r \neq s} I(\tilde{p}_{rs} \leq \alpha)}{q^2}.
\label{ucsv-dir}
\end{equation}
$CSV_U$ and $CSV_D$ represent the proportion of null hypotheses described in Section \ref{sub:procedure} that can be rejected at a given significance level $\alpha$ (after correcting for multiple testing). Unless otherwise stated, hereafter we set $\alpha = 0.05$ for the computation of the CSV indices.
By definition, $CSV_U \in [0,1]$ and $CSV_D \in [0,1]$; higher values of $CSV$ provide stronger evidence that a partition of nodes induces a community structure in a graph. 

Besides evaluating the validity of a partition as a whole with the CSV index, it may be interesting to also check whether each cluster $C_r \in \mathcal{P}_V$ is well separated from the others.
This can be done by considering, for each set $C_r$, the proportion of tests involving $C_r$ that lead to a rejection at the $\alpha$ level:
\begin{equation*}
\frac{ I(\tilde{p}_{rr} \leq \alpha) + \sum_{s \neq r} I(\tilde{p}_{rs} \leq \alpha)}{q}.
\end{equation*}

Lastly, a valuable feature of the CSV indices defined in equations \eqref{ucsv-undir} and \eqref{ucsv-dir} is that they can be directly interpreted as the proportion of null hypotheses that can be rejected for a given $\alpha$ level. However, one may additionally be interested in evaluating the ``strength'' of each rejection by weighting each rejection by the distance between $\tilde{p}_{rs}$ and $\alpha$. This can be done by considering a weighted version of CSV, which we call WCSV, where we weight each rejection $I(\tilde{p}_{rs} \leq \alpha)$ by $\frac{\alpha-\tilde{p}_{rs}}{\alpha} \in [0, 1]$. For example, for undirected graphs this yields
\begin{equation}
WCSV_U = \frac{\sum_{r=1}^q I(\tilde{p}_{rr} \leq \alpha) \frac{\alpha-\tilde{p}_{rr}}{\alpha} + \sum_{r>s} I(\tilde{p}_{rs} \leq \alpha) \frac{\alpha-\tilde{p}_{rs}}{\alpha}}{q(q+1)/2}.
\label{wcsv-undir}
\end{equation}
An equivalent definition for directed graphs can be obtained by adding weights to Equation \eqref{ucsv-dir}. Similarly to CSV, also $WCSV \in [0,1]$; moreover, $WCSV \leq CSV$. As we will show in Section \ref{sub:sim1}, CSV and WCSV typically differ for small graphs, and they tend to achieve the same value for larger graphs (Figure \ref{fig:sim1}).

\section{Comparing different network partitions and different networks}
\label{sec:netw-comp}

\subsection{Comparing different partitions of a network}\label{sub:comp-part}

Network clustering represents a common way to summarize the set of relations encoded in a network. However, in practice this task is complicated by the availability of several alternative clustering algorithms that typically produce different network partitions. When this happens, one is left with the question of which partition can separate better the retrieved clusters.

\begin{figure*}
\begin{center}
	\includegraphics[scale = 0.5]{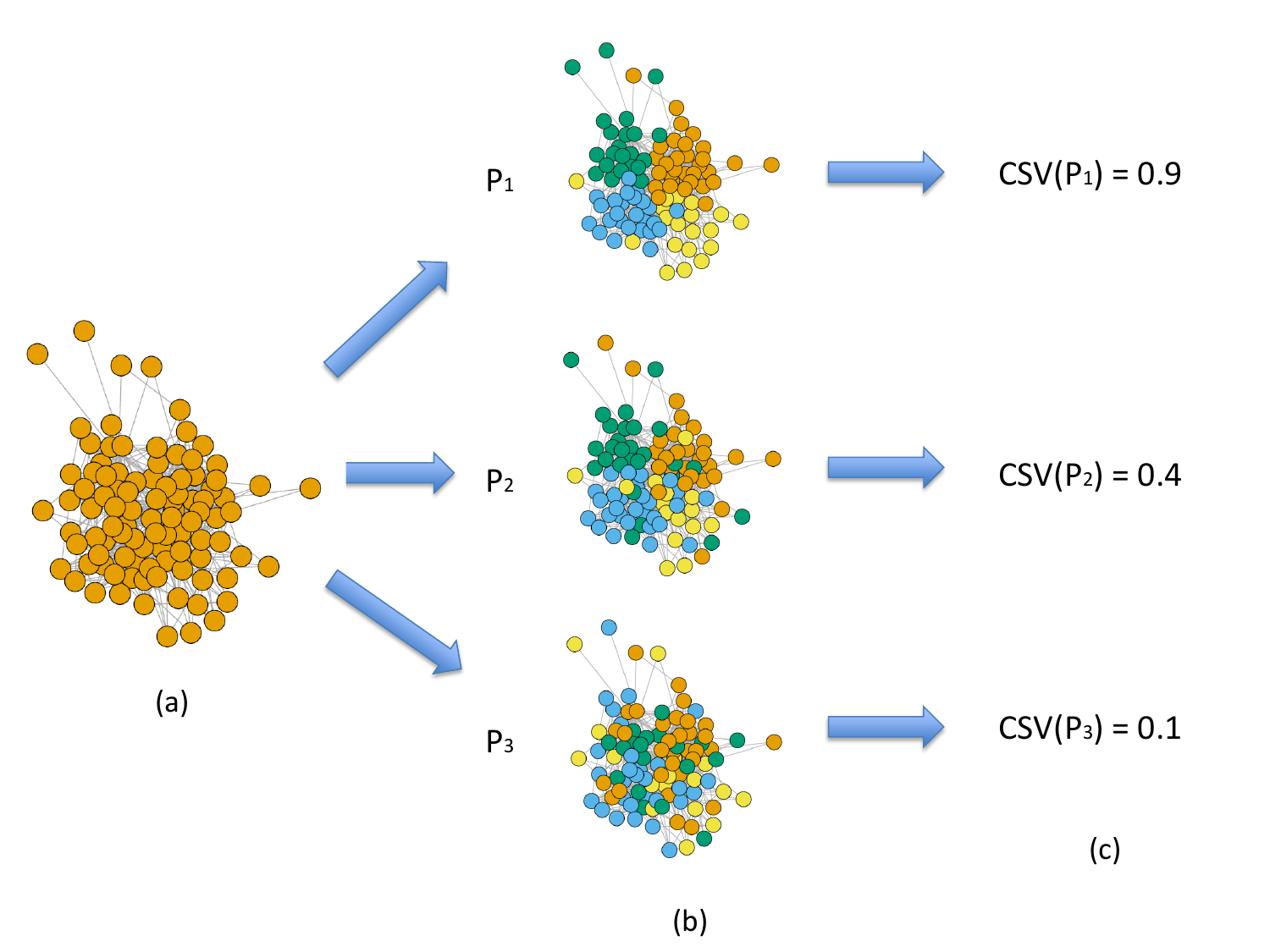}
\end{center}
	\caption{\label{fig:compare-partitions} Graphical representation of the use of the Community Structure Validation index for the comparison of different network partitions. (a) We generate a network with 100 nodes and 4 communities from the degree-corrected stochastic blockmodel illustrated in Section \ref{sub:dcsbm}. (b) We consider three different partitions of the nodes: the partition induced by true communities (top) and two partitions where 20\% (centre) and 40\% (bottom) of the nodes are assigned to the wrong cluster. (c) Computation of $CSV$ results into values close to 1 for the true communities, into intermediate values for partitions where a small proportion of the nodes are assigned to the wrong communities and into values of $CSV$ close to 0 for partitions that do not induce a community structure into the network.}
\end{figure*}

In Figure \ref{fig:compare-partitions} we show how CSV can be employed to compare different partitions of a network: by computing the value of CSV for each partition, we can obtain a quantification of how strong is the evidence that each network partition induces a community structure with well-separated communities. The higher the value of CSV, the stronger the evidence that a partition produces a clear community structure.


\subsection{Comparing community structures across networks}\label{sub:netcomp}

The difficulty to inspect and visualize relations between thousands of nodes
represents a primary challenge in the analysis of large networks. As discussed in Section \ref{sub:comp-part}, network clustering algorithms are often employed to simplify this task, summarizing a network into a set of clusters of nodes.
In addition to this, clustering algorithms are also employed to compare networks. The idea, in this context, is that we may expect similar networks to share similar communities, so that the comparison of communities in networks can point out structural similarities and differences between networks.

A simple way to put this idea in practice is to apply a clustering algorithm to the networks of interest, and check the overlap (proportion of shared genes) between the communities of each network: a high overlap between the partition $\mathcal{P}_1$ of graph  $\mathcal{G}_1$ and the partition $\mathcal{P}_2$ of graph  $\mathcal{G}_2$ can be taken as an indication that the networks share similar community structures. However, such a comparison directly compares only the allocations of nodes within $\mathcal{P}_1$ and $\mathcal{P}_2$, ignoring the different distribution of edges in the two graphs. In this Section we illustrate how CSV allows to carry out a more comprehensive assessment of the overall similarity between the community structures of two graphs.

 We propose a procedure that is based on the assessment of the validity of $\mathcal{P}_1$ as community structure for $\mathcal{G}_2$, and of $\mathcal{P}_2$ as community structure for $\mathcal{G}_1$. The idea at the basis of this approach is that if $\mathcal{G}_1$ and $\mathcal{G}_2$ have the same (or similar) community structure, then the communities extracted from one graph should also induce a community structure in the other graph.

This procedure comprises the following steps:
\begin{itemize}
\item choose a community detection method and apply it to $\mathcal{G}_1$ so as to derive its partition in $q$ communities $\mathcal{P}_1 = \{ C_{11}, ..., C_{1q} \}$. Similarly, obtain $\mathcal{P}_2$ from $\mathcal{G}_2$;
\item compute the community structure validation indices of $\mathcal{P}_1$ in $\mathcal{G}_1$ and in $\mathcal{G}_2$, and of $\mathcal{P}_2$ in $\mathcal{G}_1$ and in $\mathcal{G}_2$; 
\item compute the relative indices
\begin{equation}
\label{relative-ucsv}
R_{CSV} \left( \mathcal{P}_i|\mathcal{G}_j \right) = \frac{ CSV(\mathcal{P}_i|\mathcal{G}_j) }{ CSV(\mathcal{P}_i|\mathcal{G}_i) }, \: i \neq j \in \{ 1,2 \},
\end{equation}
which compare the values of the CSV index of partition $\mathcal{P}_i$ in graph $\mathcal{G}_j$ with the  value of CSV for $\mathcal{P}_i$ in  $\mathcal{G}_i$.
\end{itemize}

The rationale behind $R_{CSV} \left( \mathcal{P}_i|\mathcal{G}_j \right)$ is that since $\mathcal{P}_i$ is the partition in communities of $\mathcal{G}_i$, we expect $CSV(\mathcal{P}_i|\mathcal{G}_i)$ to be close to 1, or at least as high as possible for $\mathcal{G}_i$. 
The value of $CSV(\mathcal{P}_i|\mathcal{G}_j)$ will be typically smaller than $CSV(\mathcal{P}_i|\mathcal{G}_i)$: we expect it to be close to 0 if $\mathcal{P}_i$ provides a bad partition for $\mathcal{G}_j$; however, if $\mathcal{P}_i$ partitions $\mathcal{G}_j$ well, the value of $CSV(\mathcal{P}_i|\mathcal{G}_j)$ can be expected to be closer to $CSV(\mathcal{P}_i|\mathcal{G}_i)$.

As a result, we expect higher values of $R_{CSV} \left( \mathcal{P}_1|\mathcal{G}_2 \right)$ and $R_{CSV} \left( \mathcal{P}_2|\mathcal{G}_1 \right)$ when $\mathcal{G}_1$ and $\mathcal{G}_2$ share similar communities; if, on the other hand, the communities in the two graphs are different, we expect $R_{CSV} \left( \mathcal{P}_1|\mathcal{G}_2 \right)$ and $R_{CSV} \left( \mathcal{P}_2|\mathcal{G}_1 \right)$ to be close to 0. In Section \ref{sub:gamba} we provide an example of this use of the CSV index.

\section{Simulations}
\label{sec:simulations}

In this section we study the behaviour of CSV in three different simulation studies. In Section \ref{sub:dcsbm} we introduce a degree-corrected stochastic blockmodel for binary graphs that we employ to simulate the data. In Section \ref{sub:sim1} we study how graph size and modularity affect the capacity of CSV to detect a clustering of nodes as valid community structure. In Section \ref{sub:sim2} we study the behaviour of CSV with respect to increasing levels of community degradation, considering at the same time different values of modularity. Finally, in Section \ref{sub:sim3} we employ CSV to compare the performance of four different network clustering algorithms.

\subsection{A degree-corrected stochastic blockmodel for binary graphs}\label{sub:dcsbm}

The assessment of the performance of the CSV indices requires a realistic generative model of graphs with weak and strong community structures. This is tipically achieved by recurring to stochastic blockmodels \citep{holland1983}, in which the probability of observing an edge between two nodes depends on the communities they belong to. A problem with stochastic blockmodels, however, is that they are often too simple to reproduce the behaviour of real networks, mainly because they assume all nodes within a community to behave similarly. This, for example, implies that in graphs generated from such models, nodes within each community have roughly the same degree: a fact, this, that is in sharp contrast with most real networks, which feature a strong heterogeneity in the degree distribution.

To overcome this limitation, several different extensions of stochastic blockmodels have been proposed \citep{wang1987,karrer2011,signorelli2018}. Among them, \citet{karrer2011} proposed a degree-corrected stochastic blockmodel (DCSBM) for edge-weighted undirected graphs where the value of an edge between nodes $i$ and $j$ depends both on their communities $C_i$ and $C_j$, and on nodal weights $w_i$ and $w_j$: $y_{ij} \sim \text{Poi} \left( w_i w_j \lambda_{C_i \: C_j} \right)$. 

To simulate networks in our simulation studies, we employ a DCSBM for binary undirected graphs that is closely related to that of \citet{karrer2011}. We assume that the probability $\pi_{ij}$ of an edge between nodes $i$ and $j$ depends both on their communities $C_i$ and $C_j$ by means of a block-interaction parameter $\theta_{C_i \: C_j}$, and on nodal weights $w_i$ and $w_j$:
\begin{gather}
y_{ij}|i \in C_i, \: j \in C_j \sim Bern(\pi_{ij}), \nonumber\\
\pi_{ij} = min(w_i w_j \theta_{C_i \: C_j}, 1), \label{eq:dcblockmodel}
\end{gather}
where $w_i > 0 \: \forall i \in V$, $\theta_{C_i \: C_j} \in [0,1]$ and 
$$\sum_i w_i I(C_i = C_r) = n_r \:\: \forall C_i, C_j.$$
Note that the weights are defined in such a way that the average nodal weight in each community is 1; to wit, $w_i > 1$ will indicate that the expected degree of node $i$ is above the average expected degree of nodes in community $C_i$.

\subsection{Simulation 1: performance of CSV with respect to modularity and number of vertices}\label{sub:sim1}

The aim of our first simulation is to evaluate how CSV is affected by the modularity \citep{newman2004} and number of vertices of the graph. CSV relies on a significance testing procedure between each pair of groups, whose power is expected to be affected both by the size of the groups between which enrichment is tested and by the extent to which the communities are well-separated in the graph (to wit, the modularity). Therefore, we expect that CSV performs better with larger and denser networks and, for a given network size, with higher modularity and smaller number of communities.

In order to assess the performance of CSV with respect to network size and modularity, we consider four sequences of graphs with number of vertices $v \in \{100, 500, 1000, 5000\}$. For each $v$, we generate a sequence of 100 binary graphs with $p=6$ communities from the DCSBM described in Section \ref{sub:dcsbm}, where the probabilities to have an edge between nodes belonging to the same community are fixed in such a way that  $E_r(\theta_{rr}) = 0.3$ and $\theta_{rr} \in [0.22, 0.38] \; \forall r \in \{1,2,...,8\}$, and we progressively increase the probability to have an edge between nodes belonging to different communities, i.e., $\theta_{rs} \in \{0, 0.003, ..., 0.297, 0.3\} \; \forall r \neq s$. Since the $\theta_{rr}$s are fixed, increasing $\theta_{rs}$ reduces the modularity of the graphs.

\begin{figure*}\centering
\includegraphics[scale = 0.33, page = 1]{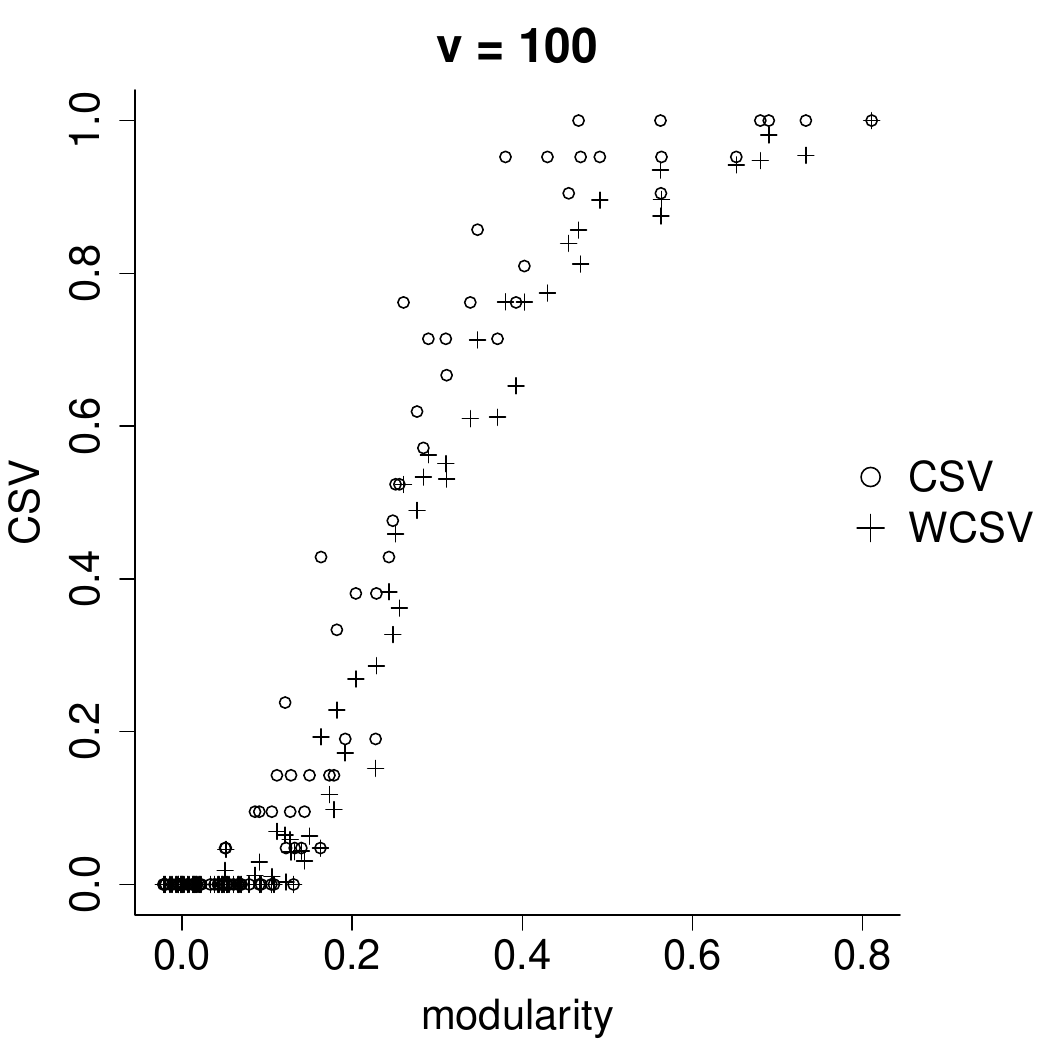}
\includegraphics[scale = 0.33, page = 2]{fig_simulation1_Heyse.pdf}
\includegraphics[scale = 0.33, page = 3]{fig_simulation1_Heyse.pdf}
\includegraphics[scale = 0.33, page = 4]{fig_simulation1_Heyse.pdf}
	\caption{\label{fig:sim1} Study of the performance of $CSV_U$ and $WCSV_U$ with respect to the number of nodes ($v$) and the modularity of the partition induced by the true communities. The capacity of CSV to validate the true communities as valid network partitions is lower in small networks (top left panel), where moderate values of CSV and WCSV may correspond to valid network partitions. With larger networks (top right, and bottom panels), instead, the true communities are easily identified as valid network partitions, provided that they generate an actual community structure (i.e., the modularity is not too low). Note how the difference between CSV and WCSV vanishes in large networks.}
\end{figure*}

For each of the graphs thus generated, we compute the $CSV_U$ and $WCSV_U$ indices associated to the partition of nodes induced by the true communities, following the procedure outlined in Sections \ref{sub:procedure} and \ref{sub:CSV}. Figure \ref{fig:sim1} shows how the value of the two indices changes for different values of modularity and $v$. We can observe that the difference between WCSV and CSV tends to vanish in large networks.
Since we are testing enrichments between the true communities, ideally we would like the CSV indices to attain their maximum possible value of 1; however, the fact that CSV relies on a hypothesis testing procedure implies that this value can only be achieved for sufficiently large networks, and for partitions that have a good level of modularity. Indeed, for small graphs ($v=100$) the index seldom achieves its maximum; this is due to the fact that if the communities are small, rejecting the null hypotheses of no enrichment is harder. For larger graphs ($v \in \{500, 1000, 5000\}$), instead, the indices achieve their maximum value when the modularity is approximately above 0.3; on the other hand, the value of the indices start to drop as the modularity of the partition generated by the true communities decreases ($Q < 0.2$). This result is desirable, because the low modularity indicates that the partition do not induce an actual community structure in the network.

\subsection{Simulation 2: behaviour of CSV with respect to community degradation}\label{sub:sim2}

In Section \ref{sub:sim1} we have assessed how network size and modularity affect the capacity of CSV to declare that the real communities result into a community structure.
The purpose of this second simulation, instead, is to understand the sensitivity of CSV to different levels of community degradation. This is important because when dealing with real data the true communities are typically be unknown, and analysts need to estimate them with a clustering algorithm that is likely to misclassify some nodes. From a practical point of view, thus, it is important to know whether CSV is capable to validate community structures even when a small proportion of nodes is misclassified.

Our desired behaviour for CSV is that a partition of nodes where most of the nodes are correctly classified, and only a small proportion of nodes is assigned to a wrong cluster, still induces a community structure in the network, and it should thus yield high CSV values. Higher proportions of wrongly classified nodes, instead, should progressively destroy the community structure, determining a sharp decrease of the CSV index.

To verify this behaviour, we generate six graphs with $v=1000$ nodes and $p = 8$ blocks from a degree-corrected stochastic blockmodel where we keep constant the probabilities of interaction within blocks, $\theta_{rr} = 0.3 \; \forall r \in \{1,2,...,8\}$. We progressively increase the probabilities of interaction between blocks $\theta_{rs}, \; r \neq s$ from 0.01 in Simulation 2A to 0.3 in Simulation 2F. Note that the modularity of the graphs decreases (from 0.68 in 2A to 0.17 in 2F) as $\theta_{rs}$ increases.

In each of the 6 scenarios considered, we take the graph thus generated and its communities as reference. Then, we generate a sequence of graphs from a degree-corrected stochastic blockmodel where we keep the same block-interaction probabilities $\theta_{rr}$ and $\theta_{rs}$, but we change community to a proportion $q$ of nodes. We consider 100 graphs for each level of community degradation $q \in \{0, 0.05, 0.10, ..., 0.95, 1\}$ and compute the CSV associated to the reference communities. 

Supplementary Figures 1 and 2 show how the distribution of CSV changes for different levels of community degradation $q$. In Supplementary Figure 1 we can observe that for high values of modularity, partitions of nodes with levels of community degradation up to 20-25\% still result into a clear community structure. The tolerance to community degradation is instead lower when the modularity decreases, as shown in Supplementary Figure 2. In both cases, the CSV index is stable around 1 for moderate values of community degradation, after which it rapidly decreases towards 0, indicating that higher levels of perturbation of the real communities break the community structure. These results show that CSV adheres to the desired behaviour described at the beginning of this section.

\section{Applications}\label{sec:application}

In this Section we discuss two examples that illustrate how community structure validation indices can be employed to compare different partitions of the same network (Section \ref{sub:cospons}) and to compare community structures across networks (Section \ref{sub:gamba}). We begin with an application to a network of collaborations between deputies in the Italian Chamber of Deputies, where we explore the usefulness of incorporating metadata in community detection algorithms, and we conclude with a comparison of community structures across 30 tissue-specific gene co-regulation networks.

\subsection{Community detection and metadata: an application to bill cosponsorships}\label{sub:cospons}

In Section \ref{sec:intro} we mentioned that the use of nodal metadata for community detection and community validation has recently become an active topic of discussion among network scientists. Network data are often accompanied by annotations, or \textit{metadata}, that describe properties of nodes (such as age, gender and ethnicity in a social network, or data capacity and location of nodes in the Internet network). \citet{newman2016} argued in favour of the use of metadata for community detection, and proposed a community detection algorithm that uses both the graph structure and one categorical variable to identify clusters of nodes. They claimed that when a network is not very informative about communities, a given set of nodal labels (metadata) can improve the accuracy of the clustering. On the other hand, \citet{peel2017} observed that whereas metadata describe the features of a node, the concept of community is about the edges that exist between the nodes, rather than about the nodes themselves, and they warned against the common practice of using metadata to validate communities in networks. 

In this Section we use the CSV index to compare the performance of the 4 community detection algorithms considered in Section \ref{sub:sim3}, which do not exploit metadata, to that of the method from \citet{newman2016} (referred to as \textit{``Newclau''} hereafter), which makes use of metadata. The network that we use for this comparison is a bill cosponsorship network that summarizes the bill cosponsorship activity of 663 Deputies in Italian Chamber of Deputies during the XVI legislature (2008-2013) \citep{briatte2016, signorelli2018}.
In particular, we consider a binary, undirected network where each Deputy is a node, and an edge indicates that two Deputies have cosponsored together at least one bill during the legislature; moreover, we consider gender and party affiliation as metadata. A graphical illustration of the network is presented in Supplementary Figure 4; it can be observed that nodes therein tend to cluster according to party affiliation, but not based on gender.

Application of the fast greedy clustering algorithm leads to the identification of 3 clusters, whereas Louvain and the leading eigenvalue methods detect 4 clusters; finally, walktrap extracts 4 clusters and a few isolate nodes. Since the Newclau method does not select automatically the number of communities $K$, we set $K = 3$ for its comparison to fast greedy, and $K = 4$ to compare it to Louvain, leading eigenvalue and walktrap.
We compare each of the partitions thus obtained following the procedure for the comparison of different partitions of the same network outlined in Section \ref{sub:comp-part}.

Table \ref{tab:ari3} reports the value of the adjusted random index (ARI) for the methods that produce 3 clusters; it can be observed that Newclau using party or gender as metadata result in almost the same partition, which is instead quite different from that of fast greedy. The values of the CSV index, reported in Table \ref{tab:cvs-cl3}, indicate that the partitions identified by the Newclau method induce somewhat better community structures than fast greedy.

\begin{center}
\begin{table}[h!]
\begin{tabular}{cccc}
\hline
 Method & fg & np3 & nc3 \\\hline
 Fast greedy (fg) & 1 & 0.6 & 0.6\\
 Newclau + party, 3 clusters (np3) & 0.6 & 1 & 0.992\\
 Newclau + gender, 3 clusters (nc3) & 0.6 & 0.992 & 1\\
 \hline
\end{tabular}
\caption{Value of the ARI for network partitions with 3 clusters.}
\label{tab:ari3}
\end{table}
\end{center}

\begin{center}
\begin{table}[h!]
\begin{tabular}{cc}
\hline
 Method & CSV index\\\hline
 Fast greedy & 0.667\\
 Newclau + party, 3 clusters & 0.833\\
 Newclau + gender, 3 clusters & 0.833\\
 \hline
\end{tabular}
\caption{Value of the CSV index for network partitions with 3 clusters.}
\label{tab:cvs-cl3}
\end{table}
\end{center}

As concerns the comparison of the methods that produce 4 clusters, the values of the ARI presented in Table \ref{tab:ari4} show that three methods (Louvain, Newclau using party as metadata, and Newclau using gender as metadata) yield very similar network partitions, whereas walktrap and leading eigenvalue retrieve somewhat different partitions. The values of the CSV index, reported in Table \ref{tab:cvs-cl4}, indicate that walktrap, Louvain and Newclau retrieve partitions that result into a clear community structure within the cosponsorship network, outperforming the leading eigenvalue method.

\begin{center}
\begin{table*}[h!]
\begin{tabular}{cccccc}
\hline
 Method & wt & le & lou & np4 & nc4 \\\hline
 Walktrap (wt) & 1 & 0.767 & 0.903 & 0.894 & 0.897\\
 Leading eigenvalue (le) & 0.767 & 1 & 0.831 & 0.825 & 0.834\\
 Louvain (lou) & 0.903 & 0.831 & 1 & \textbf{0.970} & \textbf{0.982}\\
 Newclau + party, 4 clusters (np4) & 0.894 & 0.834 & \textbf{0.970} & 1 & \textbf{0.988}\\
 Newclau + gender, 4 clusters (nc4) & 0.897 & 0.834 & \textbf{0.982} & \textbf{0.988} & 1\\
 \hline
\end{tabular}
\caption{Value of the ARI for network partitions with 4 clusters.}
\label{tab:ari4}
\end{table*}
\end{center}

\begin{center}
\begin{table}[h!]
\begin{tabular}{cc}
\hline
Method & CSV index\\\hline
Walktrap & 0.90\\
Leading eigenvalue & 0.70\\
Louvain & 0.90\\
 Newclau + party, 4 clusters & 0.90\\
 Newclau + gender, 4 clusters & 0.90\\
 \hline
\end{tabular}
\caption{Value of the CSV index for the partitions with 4 clusters.}
\label{tab:cvs-cl4}
\end{table}
\end{center}

\subsection{Comparison of 30 tissue-specific gene regulatory networks}\label{sub:gamba}

In this section we apply the procedure for the comparison of community structures in different networks outlined in Section \ref{sub:netcomp} to a collection of 30 tissue specific gene co-regulatory networks that were inferred in \citet{gambardella2013}. The data are publically available from \url{https://bitbucket.org/ggambard/dina-differential-network-analysis}.

\begin{figure*}
\begin{center}
	\includegraphics[scale = 0.6, page=2]{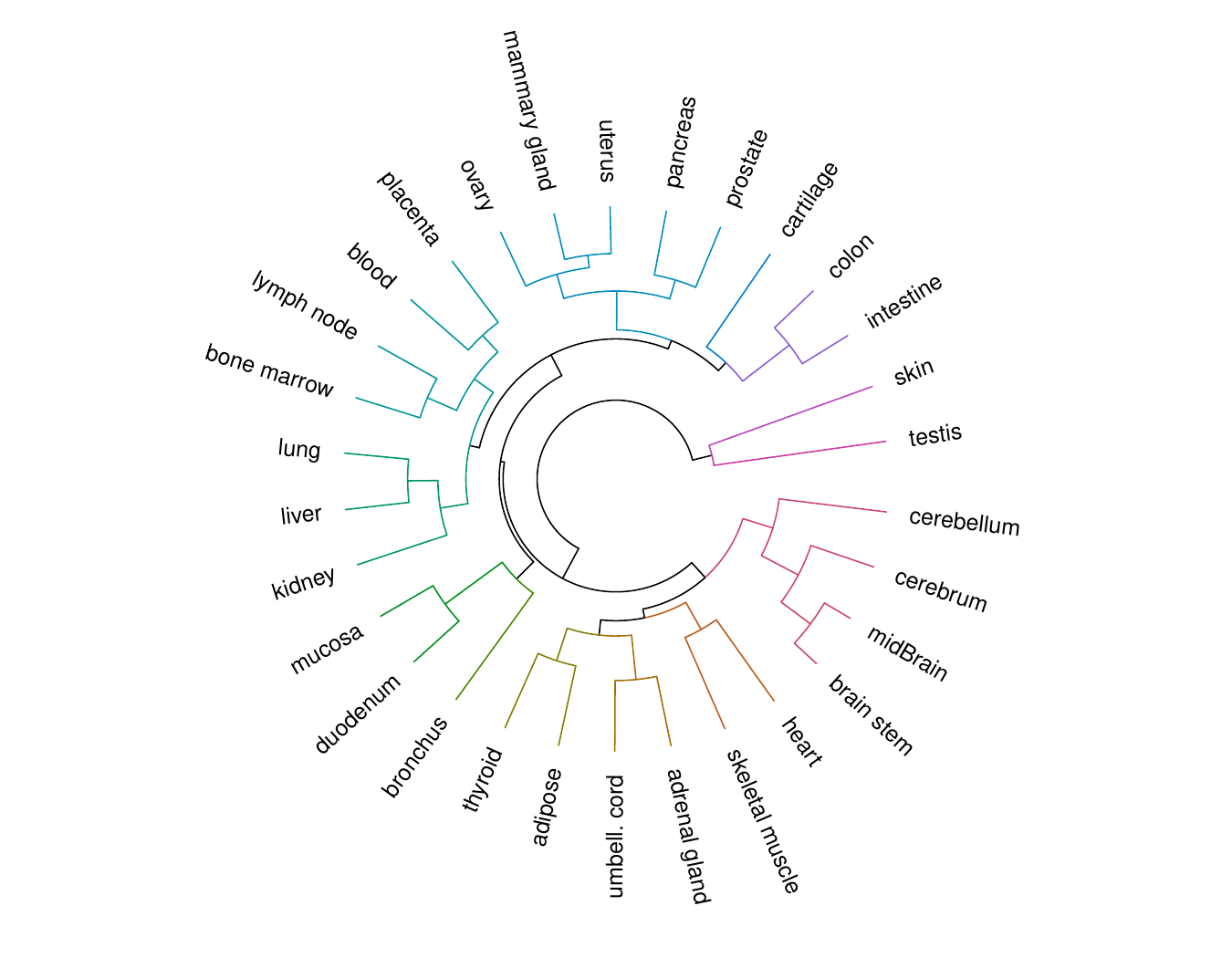}
\end{center}
	\caption{\label{fig:dendogram} Dendrogram associated to the complete-linkage clustering of 30 tissue-specific gene co-regulation networks. The resulting 13 clusters of tissues are highlighted in different colours.}
\end{figure*}

In this study, 30 tissue-specific co-regulation networks comprising 11730 human genes were reverse-engineered from data from 2930 microarrays. The identification of similarities and differences between these networks is complicated both by the size of the networks, which prevents an effective graphical comparison, and by the large number of networks considered. Here, we use the $CSV_1$ index to compare community structures between each pair of networks as illustrated in Section \ref{sub:netcomp}, and derive a similarity matrix between networks that will provide a synthetic overview of the networks at hand. We apply the Louvain network clustering method \citep{blondel2008} to obtain a partition $\mathcal{P}_i$ of graph $\mathcal{G}_i$ and a partition $\mathcal{P}_j$ of $\mathcal{G}_j$. To guarantee statistical power in the testing procedure, only communities of size greater than 5 are retained in our analysis. Then, we compute the relative indices $R_{CSV}(\mathcal{P}_i|\mathcal{G}_j)$ defined in Section \ref{sub:CSV} for each pair of graphs, deriving a $30 \times 30$  matrix  $R$ such that $R_{ij}=R_{CSV}(\mathcal{P}_i|\mathcal{G}_j)$. In order to have a general picture of the tissues similarity, we build a similarity matrix  $S= \left( R + R^T \right)/2$, derive the corresponding distance matrix $D = 1 - S$ and apply a complete-linkage clustering over $D$. The dendrogram resulting from the clustering is represented in Figure \ref{fig:dendogram} in circular layout. By cutting the dendrogram at different heights, one can partition the 30 tissue-specific networks into a different number of clusters. In Figure \ref{fig:dendogram}, in particular, we discriminate among 13 clusters, highlighted in different colours. Three of these clusters are, in fact, singletons (testis, skin and cartilage).

Ideally, one could expect gene networks associated to tissues that share similar structure or function to be more similar with each other and, thus, to obtain as a result higher $R_{CSV}$s and similarities for such tissues so that, in the end, most of the clusters displayed in the dendrogram would reflect analogies in tissues' structure and function. Indeed, this intuition seems corroborated (at least in part) by the results in Figure \ref{fig:dendogram}: for example, one cluster exclusively comprises all the cerebral tissues considered in the analysis (cerebrum, cerebellum, mid brain and brain stem), and one the only two striated muscles (heart and skeletal muscle) involved in the study. Moreover, the reproductive system female organs (mammary gland, uterus and ovary) are linked together in the same cluster, and the two tissues from the lower digestive system (colon and intestine) form together a unique cluster. Although a biological interpretation of these results is beyond the scope of this paper, overall we can observe that CSV tends to find higher similarities between graphs corresponding to tissues with structural or functional similarities, in line with our initial expectations.

\section{Conclusion}
\label{sec:concl}

Community structure is a commonly observed property of real networks. The term refers to the presence, in a network, of groups of nodes (also referred to as modules or communities) that are strongly tied to each other, and sporadically connected to other nodes in the network; this feature is often exploited to simplify the interpretation of large networks and to identify their relevant modules. Whereas the problem of community detection in networks has received wide attention, the assessment of the validity of a partition of nodes as community structure for a given graph has received considerably less attention. In this article, we have proposed a strategy to perform community structure validation of a partition $\mathcal{P}$ of nodes that consists of two steps. First, the presence of enrichment between any two sets in $\mathcal{P}$ is assessed with a one-tailed modification of NEAT, the test for network enrichment analysis proposed by \citet{signorelli2016}. Then, the results from these tests are summarized into a synthetic index for community structure validation (CSV), which can either be unweighted (CSV) or weighted (WCSV).

The rationale behind the CSV indices, which range between 0 and 1, is that they will approach 1 when there is evidence of a strong separation between the sets in $\mathcal{P}$ - to wit, when $\mathcal{P}$ induces a clear community structure - and they will be close to 0 otherwise. In this sense, the CSV indices can be used to validate $\mathcal{P}$ as community structure.

Our simulations indicate that the performance of the proposed indices is poor for very small networks (e.g., $v = 100$), where the hypothesis testing procedure is not enough powerful to reject the null hypothesis of no enrichment between gene sets, but it heavily improves for larger networks ($v \geq 500$), where CSV behaves as expected and the difference between $CSV$ and $WCSV$ rapidly vanishes (Section \ref{sub:sim1}).
Thus, CSV is capable to identify whether a partition of nodes induces a community structure as long as the network at hand is not too small. It is also robust to a moderate extent of community degradation (Section \ref{sub:sim2}), thus making allowance for the possibility that a clustering algorithm may misclassify a small number of nodes. In Section \ref{sub:sim3}, we have employed CSV to compare four popular clustering algorithms for networks on synthetic data. Our results indicate that the Louvain and walktrap clustering algorithms typically outperform the leading eigenvalue and fast greedy methods.

As illustrated in Section \ref{sec:netw-comp}, CSV indices can be employed to compare the goodness of different partitions of the same network as community structures, as well as to evaluate whether different networks share similar or different community structures. An example of the first task is given in Section \ref{sub:cospons}, where we have compared partitions obtained with several different clustering algorithms in a network representing cosponsorship between Deputies in the Italian Parliament. An example of the second task, instead, is provided in Section \ref{sub:gamba}, where we have employed community structure validation to quantify the extent of similarity across the community structures of 30 tissue-specific gene co-regulation networks.

A limitation of the proposed indices is that because they depend on an hypothesis testing procedure, they are affected by a lack of power for very small or very sparse networks, and they may be ``too powerful'' when very large and dense networks are considered instead. Moreover, the choice of the significance level $\alpha$ is somehow arbitrary. Although throughout this paper we kept $\alpha = 0.05$ fixed for illustration purposes, it may be advisable to consider a larger $\alpha$ when small networks are at hand, and smaller ones with larger networks.

\section*{Declarations}

\subsection*{Funding}
The authors gratefully acknowledge funding from the COST Action CA15109 ``European Cooperation for Statistics of Network Data Science'', supported by COST (European Cooperation in Science and Technology). The work of Luisa Cutillo has been supported by the European Union under Horizon 2020, Marie Sklodowska-Curie Individual Fellowship CONTESSA (ID: 660388).

\subsection*{Conflicts of interest/Competing interests}
The authors declare that they have no conflict of interest.

\subsection*{Availability of data and material}
An online Supplementary Material is available from the journal's website.

\subsection*{Code availability}
The \texttt{R} code used to implement the CSV approach, for the simulations and the data analyses is available at \href{https://github.com/mirkosignorelli/csv}{\texttt{github.com/mirkosignorelli/csv}}.

\subsection*{Authors' contributions}
The authors contributed equally to this study.

\subsection*{Acknowledgements}
The authors would like to thank Diego di Bernardo for his valuable comments on the application illustrated in Section \ref{sub:gamba}.

\bibliographystyle{apa}
\bibliography{bibliography}

\end{document}